\documentstyle[seceq,preprint]{ptptex}

\newcommand{\bra}[1]{\langle {#1} |}     
\newcommand{\ket}[1]{| {#1} \rangle}     
\newcommand{\bbra}[1]{\langle\!\langle {#1} |}     
\newcommand{\kket}[1]{| {#1} \rangle\!\rangle}     
\newcommand{\rket}[1]{| {#1} )}     

\title{
Deformed Boson Scheme in Time-Dependent Variational Method. I
}
\subtitle{The Case of Many-Body System Constituted of One 
Kind Boson Operator}    

\author{
Atsushi {\sc Kuriyama}, 
Constan\c{c}a {\sc Provid\^encia}$^{*}$, \\
Jo\~ao da {\sc Provid\^encia}$^{*}$, Yasuhiko {\sc Tsue}$^{**}$ 
and Masatoshi {\sc Yamamura}
}

\inst{
Faculty of Engineering, Kansai University, Suita 564-8680, Japan\\
$^{*}$Departamento de Fisica, Universidade de Coimbra, P-3000 Coimbra, 
Portugal\\
$^{**}$Physics Division, Faculty of Science, Kochi University, Kochi 780-8520, 
Japan
}

\recdate{
}

\abst{
Various aspects of the deformed boson scheme are investigated 
with aiming of applying to the time-dependent variational method 
for many-boson system. The case of the multiboson states is 
also discussed. In analogy with the time-dependent Hartree-Fock 
theory in many-fermion system, the classical aspect of the 
deformed boson is discussed. 
}

\begin{document}

\maketitle

\section{Introduction}

The time-dependent variational method has played a central 
role in the studies of many-body systems. One of the typical 
examples can be found in the time-dependent Hartree-Fock theory 
in canonical form.\cite{MMSK}
This theory starts with a Slater determinant parametrized by 
complex parameters, we can describe the time-evolution of 
many-fermion system under a certain condition, which we call 
the canonicity condition, they can be expressed in terms of 
the canonical variables of classical mechanics and the time-evolution 
of the system is reduced to solving the Hamilton's equation of 
motion. In addition to the case of many-fermion system, 
we applied the above-mentioned idea to the time-evolution 
of many-boson system in terms of the coherent and the squeezed 
state. This application has been reviewed in Ref.\citen{KPTYs}. 

For the time-dependent variational method, selection of the trial 
function for the variation is the first task. 
In the case of many-boson system, the simplest may be 
so called boson coherent state. However, the boson coherent 
state is not always a unique trial function. 
Depending on the Hamiltonian under investigation, the selection 
should be done. In relation to this task, we proposed a possible 
generalization from the conventional boson coherent 
state.\cite{KPTY3} 
With the help of this generalization, we described a boson 
system interacting with an external harmonic oscillator 
in terms of the damped and the amplified motion.\cite{KPTY4} 
Another generalization by using an appropriate mapping method 
was carried out\cite{CP} by the two of the present 
authors (C. P \& J. da P.) with Brito in order to investigate 
the $su(2)$-Lipkin model in the point of view of classical 
$q$-deformation of $su(2)$- and $Os(1)$-algebra. 
Recently, an interesting idea was presented by 
Penson and Solomon.\cite{Penson} 
Under a certain theoretical framework, they found a possible 
generalization from the conventional boson coherent state. 
Especially, they investigated their form from the viewpoint 
of the deformed boson scheme. In this scheme, the function 
$[x]_q$ plays a central role. They mentioned that for historical 
reason some specific form of $[x]_q$ gained a particular 
popularity,\cite{Mac} 
while another deformed boson scheme is possible and 
was really introduced earlier.\cite{AC} 
After stressing that there exists an 
infinite number of possible deformations, they proposed 
a specific form for $[x]_q$. 
By changing the value of the parameter specifying the deformation, 
this function gives us various types of the superposition 
of many-boson system, which are different from that given 
in the most popular form. 
We would like to discuss
the generalization of the boson coherent state 
in the framework of  the deformed boson scheme. 
With the aid of this viewpoint, the treatment of the 
time-dependent variational method in many-boson system is 
expected to become more transparent than the level shown in 
Ref.\citen{KPTY3}, where no reference was given to 
the deformed boson scheme. 

The main aim of this series of the papers is to investigate 
extensively the generalization of the boson coherent state 
which has been used in our variational description 
of many-boson system in the framework of the deformed 
boson scheme. In Part (I), the present paper, we treat the case 
of one kind of boson operator. By introducing a reasonable function 
for the boson number, the conventional boson coherent state 
is generalized. Further, we can find an operator, the eigenstate of 
which is just the above state. This operator is nothing but the 
deformed boson. For the demonstration of the powerfulness of 
the generalization, three concrete examples are discussed ; 
(1) the most popular form, (2) the form given in Ref.\citen{Penson} 
and (3) the combined form of (1) and (2). The idea for the multiboson 
states is presented in a way different from that 
in Ref.\citen{Penson} ; the use of the MYT boson mapping.\cite{MYT} 
The motivation comes from the private communication by 
Karpeshin, J. da Provid\^encia and C. Provid\^encia for 
the problem on anharmonic effects in large amplitude 
vibration of metal clusters.\cite{KPP} 
Our deformed boson scheme is closely connected with classical 
mechanics and the connection is discussed in the framework of the 
canonicity condition which appears in the time-dependent 
Hartree-Fock theory.\cite{MMSK}

In the next section, the basic idea of our treatment will 
be mentioned. Section 3 will be devoted to discussing 
concrete examples for the deformations. 
In \S 4, the idea for the multiboson states will be described. 
In \S 5, classical aspect of the deformed boson scheme 
will be mentioned and in \S 6, short concluding remarks will be given.

\section{Basic scheme}

We set out by investigating a space constructed 
by the boson operator $({\hat c} , {\hat c}^*)$ : 
\begin{equation}\label{2-1}
[ {\hat c} , {\hat c}^* ]=1 \ .
\end{equation}
This boson space is spanned by the basis $\{ \ket{n} \}$ given 
by the form 
\begin{subequations}\label{2-2}
\begin{eqnarray}
& &\ket{n}=(\sqrt{n!})^{-1} ({\hat c}^*)^n \ket{0} \ , \quad
(n=0, 1, 2, \cdots)\label{2-2a}\\
& &{\hat c}\ket{0}=0 \ . \label{2-2b}
\end{eqnarray}
\end{subequations}
For the operators ${\hat c}$ and ${\hat c}^*$, we have the relations 
\begin{subequations}\label{2-3}
\begin{eqnarray}
& &{\hat c}^*{\hat c}={\hat N} \ , \label{2-3a}\\
& &{\hat c}{\hat c}^*={\hat N}+1 \ , \label{2-3b}
\end{eqnarray}
\end{subequations}
\begin{eqnarray}
& &{\hat c}\ket{n}=\sqrt{n}\ket{n-1} \ , \qquad
{\hat c}^*\ket{n}=\sqrt{n+1}\ket{n+1} \ , \label{2-4}\\
& &[ {\hat N} , {\hat c} ]=-{\hat c} \ , \qquad
[ {\hat N} , {\hat c}^* ]={\hat c}^* \ , \label{2-5}\\
& &{\hat N} \ket{n} = n \ket{n} \ . \label{2-6}
\end{eqnarray}
The relation (\ref{2-3a}) is the definition of the operator 
${\hat N}$ and, with the use of the relations (\ref{2-1}), 
(\ref{2-2a}) and (\ref{2-3a}), the others are derived. 

In this space, we introduce the following wave packet : 
\begin{subequations}\label{2-7}
\begin{eqnarray}
& &\ket{c}=(\sqrt{\Gamma})^{-1}
\sum_{n=0}^\infty f(n) (\sqrt{n!})^{-1} \gamma^n \ket{n} \ . \quad
(\langle c \ket{c}=1 ) \label{2-7a}\\
& &\Gamma=\sum_{n=0}^\infty f(n)^2 (n!)^{-1} (|\gamma|^2)^n \ .
\label{2-7b}
\end{eqnarray}
Here, $(\gamma, \gamma^*)$ denotes complex parameter. 
The quantity $f(n)$ is a well-behaved function of $n$ obeying 
\begin{equation}\label{2-7c}
f(n)=1 \quad {\hbox{\rm for}}\quad n=0, 1 \ , 
\qquad f(n)>0 \quad {\hbox{\rm for}}\quad 
n=2, 3, \cdots \ . 
\end{equation}
\end{subequations}
If $f(n)=1$, the wave packet $\ket{c}$ is nothing but a conventional 
boson coherent state. 
We can see that the infinite series (\ref{2-7b}) is convergent  
in the domain $0 \leq |\gamma|^2 < \infty$. 
In Ref.\citen{KPTY3}, we started with the state 
\begin{equation}\label{2-8}
\ket{c}=\left(\sqrt{\Gamma}\right)^{-1}
\exp \left( \gamma {\hat c}^* {\tilde f}({\hat N})\right)\ket{0} \ .
\end{equation}
Here, ${\tilde f}({\hat N})$ is a function of ${\hat N}$, 
the inverse of which is definable. The function $f(n)$ 
in the relation (\ref{2-7a}) is related with ${\tilde f}(n)$ 
through 
\begin{eqnarray}\label{2-9}
& &f(n)={\tilde f}(0){\tilde f}(1)\cdots {\tilde f}(n-1) \ , \qquad
(n=1, 2, 3, \cdots) \nonumber\\
& &f(0)=1 \ .
\end{eqnarray}
Therefore, the state (\ref{2-8}) is essentially the 
same as the state (\ref{2-7a}). 
Our main interest is in applying the wave packet $\ket{c}$ 
to the time-dependent variational method as a trial state : 
\begin{equation}\label{2-10}
\delta \int \bra{c} i\partial_t - {\hat H} \ket{c} dt=0 \ .
\end{equation}
Here, ${\hat H}$ denotes a Hamiltonian expressed in terms of 
$({\hat c} , {\hat c}^*)$ and the variation is performed 
through the parameter $(\gamma, \gamma^*)$. 

We note that the state (\ref{2-7a}) can be rewritten in the 
following form : 
\begin{subequations}\label{2-11}
\begin{eqnarray}
& &\ket{c}=\sqrt{\Gamma_0/\Gamma}\cdot f({\hat N})\ket{c^0} \ , 
\label{2-11a}\\
& &\ket{c^0}=\left(\sqrt{\Gamma_0}\right)^{-1}
\exp(\gamma{\hat c}^*)\ket{0} \ , \qquad 
\Gamma_0=\exp (|\gamma|^2) \ . \label{2-11b}
\end{eqnarray}
\end{subequations}
In relation to the wave packet $\ket{c}$, we introduce the operators 
defined by 
\begin{equation}\label{2-12}
{\hat \gamma}=f({\hat N}){\hat c} f({\hat N})^{-1} \ , \qquad
{\hat \gamma}^*=f({\hat N})^{-1}{\hat c}^* f({\hat N}) \ .
\end{equation}
For the sake of the property (\ref{2-7c}), $f({\hat N})^{-1}$ 
can be defined. The form (\ref{2-11}) and the definition (\ref{2-12}) 
give us the relation 
\begin{equation}\label{2-13}
{\hat \gamma}\ket{c}=\gamma\ket{c} \ . 
\end{equation}
Here, we used ${\hat c}\ket{c^0}
=\gamma\ket{c^0}$. 
The relation (\ref{2-13}) shows that the state $\ket{c}$ is, 
in some sense, a possible generalization from the conventional 
boson coherent state. 
With the use of ${\hat \gamma}^*$, the state $\ket{n}$ can be 
expressed in the form 
\begin{subequations}\label{2-14}
\begin{eqnarray}
& &\rket{n}=(\sqrt{f(n)^{-2} n!})^{-1} ({\hat \gamma}^*)^n \rket{0} \ , 
\quad (n=0, 1, 2, \cdots ) \label{2-14a}\\
& &{\hat \gamma}\rket{0}=0 \ . \label{2-14b}
\end{eqnarray}
\end{subequations}
Of course, $\rket{n}$ is identical to $\ket{n}$. 
Further, we have the relations 
\begin{subequations}\label{2-15}
\begin{eqnarray}
& &{\hat \gamma}^*{\hat \gamma}={\hat N} f({\hat N})^{-2}
f({\hat N}-1)^2  \ , \qquad\qquad\qquad \label{2-15a}\\
& &{\hat \gamma}{\hat \gamma}^*=({\hat N}+1)f({\hat N}+1)^{-2}
f({\hat N})^2 \ , \label{2-15b}
\end{eqnarray}
\end{subequations}
\begin{eqnarray}
& &{\hat \gamma}\rket{n}=\sqrt{nf(n)^{-2} f(n-1)^2}\rket{n-1} \ , 
\nonumber\\
& &{\hat \gamma}^*\rket{n}=\sqrt{(n+1)f(n+1)^{-2}f(n)^2}\rket{n+1} \ .
\label{2-16}\\
& &[ {\hat N} , {\hat \gamma} ]=-{\hat \gamma} \ , \qquad
[ {\hat N} , {\hat \gamma}^* ]={\hat \gamma}^* \ , \label{2-17}\\
& &{\hat N}\rket{n} = n \rket{n} \ .\label{2-18}
\end{eqnarray}
The above relations can be interpreted in terms of the 
deformed boson scheme. 
We define the function $[x]_q$ in the form 
\begin{eqnarray}\label{2-19}
& &[x]_q=x f(x)^{-2} f(x-1)^2 \ , \quad (x=n, {\hat N}\ , \quad 
n\ge 1) 
\nonumber\\
& &[0]_q=0 \ .
\end{eqnarray}
Then, $[n]_q!$ is given by 
\begin{eqnarray}\label{2-20}
[n]_q!&=&n f(n)^{-2}f(n-1)^2\cdot (n-1)f(n-1)^{-2}f(n-2)^2\cdots
1\cdot f(1)^{-2}f(0)^2 \nonumber\\
&=&f(n)^{-2} n! \ , \qquad\qquad (n\ge 1) \  \nonumber\\
{[0]}{}_q{!}&=&1 \ .
\end{eqnarray}
For $n=0, 1$, the relation (\ref{2-20}) gives us 
$[0]_q!=[1]_q!=1$. 
Thus, the relation (\ref{2-14})$\sim$(\ref{2-18}) are summarized 
in the framework of the deformed boson scheme. 
Of course, $({\hat \gamma}, {\hat \gamma}^*)$ denotes the deformed 
boson. The relations (\ref{2-15}) and (\ref{2-19}) give us the 
following commutation relation : 
\begin{eqnarray}\label{2-21}
[ {\hat \gamma} , {\hat \gamma}^* ] 
&=& [{\hat N}+1]_q - [{\hat N}]_q \nonumber\\
&=& \Delta_{\hat N}({\hat \gamma}^*{\hat \gamma}) \ .
\end{eqnarray}
Here, $\Delta_{\hat N}({\hat \gamma}^*{\hat \gamma})$ 
denotes the difference with $\Delta{\hat N}=1$. 
Concerning the relation between ${\hat \gamma}{\hat \gamma}^*$ 
and ${\hat \gamma}^*{\hat \gamma}$, there exist infinite 
possibilities such as shown in the form 
\begin{eqnarray}\label{2-22}
& &{\hat \gamma}{\hat \gamma}^*-[f({\hat N})^4f({\hat N}+1)^{-2}
f({\hat N}-1)^{-2}+F({\hat N})]{\hat \gamma}^*{\hat \gamma} 
\nonumber\\
&=&f({\hat N}+1)^{-2}f({\hat N})^2-F({\hat N}){\hat N}f({\hat N})^{-2}
f({\hat N}-1)^2 \ .
\end{eqnarray}
The form depends on the choice of the function $F({\hat N})$.

\section{Concrete examples}

In this section, we will discuss some concrete examples for 
the choice of the function $f(x)\ (x=n, {\hat N})$. 

\subsection{The most popular form}

This case starts in the well known form 
\begin{equation}\label{3-1}
[x]_q=(q^x-q^{-x})/(q-q^{-1}) \ , \quad (x=n, {\hat N})
\end{equation}
where we assume that $q$ is a real number in the domain 
$0<q\leq 1$.\cite{Mac} 
The function $f(x)$ can be determined through the relation 
(\ref{2-19}) : 
\begin{equation}\label{3-2}
x f(x)^{-2} f(x-1)^2 = (q^x-q^{-x})/(q-q^{-1}) \ .
\end{equation}
Since $x=n\ (n=0, 1, 2, \cdots)$, the relation (\ref{3-2}) 
gives us the following recursion formula : 
\begin{equation}\label{3-3}
f(n)=\sqrt{n(q-q^{-1})/(q^n-q^{-n})} \ f(n-1) \ .
\end{equation}
By solving the above relation successively, we have 
\begin{eqnarray}\label{3-4}
f(n)&=&
\cases{ 1\ , \quad (n=0, 1) \cr
        \displaystyle 
        \sqrt{n! \prod_{k=2}^n (q-q^{-1})/(q^k-q^{-k})} 
=\sqrt{n!/\prod_{k=2}^n\left(\sum_{m=-(k-1)/2}^{(k-1)/2}
(q^2)^m\right)} \ .}\nonumber\\
& &\qquad\qquad\qquad\qquad\qquad\qquad\qquad\qquad\qquad 
(n=2, 3, 4, \cdots)
\end{eqnarray}
Of course, $f({\hat N})$ is obtained by replacing $n$ 
with ${\hat N}$. As $f(n)$, if we choose the form 
(\ref{3-4}), our scheme is reduced to the most popular form. 
Since $f(n)$ is symmetric with respect to $q^1$ and $q^{-1}$, 
it is enough to investigate the case in the domain 
$0<q\le 1$. 
For $n=2, 3, 4, \cdots$, two special cases of the form 
(\ref{3-4}) are as follows : 
\begin{subequations}\label{3-5}
\begin{eqnarray}
& &f(n) \rightarrow 0 \qquad {\hbox{\rm for}}\qquad 
q\rightarrow 0 \ , \label{3-5a}\\
& &f(n) = 1 \qquad  {\hbox{\rm for}}\qquad 
q=1 \ . \label{3-5b}
\end{eqnarray}
\end{subequations}
In these cases, the state $\ket{c}$ can be expressed as 
\begin{subequations}\label{3-6}
\begin{eqnarray}
\ket{c}&\rightarrow&
(\sqrt{\Gamma})^{-1}(\ket{0}+\gamma\ket{1}) \nonumber\\
&=&(1+|\gamma|^2)^{-1/2}\cdot (1+\gamma{\hat c}^*)\ket{0} 
\quad {\hbox{\rm for}}\quad q\rightarrow 0 \ , \label{3-6a}\\
\ket{c}&\rightarrow&
(\sqrt{\Gamma})^{-1}\sum_{n=0}^\infty (\sqrt{n!})^{-1}
\gamma^n\ket{n} \nonumber\\
&=&\exp(-|\gamma|^2/2)\cdot \exp(\gamma{\hat c}^*)\ket{0} 
\ (=\ket{c^0})
\quad {\hbox{\rm for}}\quad q=1 \ . \label{3-6b}
\end{eqnarray}
\end{subequations}
The state (\ref{3-6a}) is  the simplest mixture and the state 
(\ref{3-6b}) is nothing but the conventional boson coherent 
state. 
Therefore, the state $\ket{c}$ in the general case is an 
intermediate mixture between the above two cases. 
We have the relation familiar to the deformed boson scheme 
\begin{equation}\label{3-7}
{\hat \gamma}{\hat \gamma}^*-q^{-1}{\hat \gamma}^*{\hat \gamma}
=q^{{\hat N}} \ , 
\end{equation}
if $F({\hat N})$ in the relation (\ref{2-22}) is 
chosen as 
\begin{equation}\label{3-8}
F({\hat N})=q^{-1}-f({\hat N})^4 f({\hat N}+1)^{-2}
f({\hat N}-1)^{-2} \ .
\end{equation}

\subsection{The form presented by Penson and Solomon}

Our scheme is reduced to the form proposed by 
Penson and Solomon,\cite{Penson}
if $f(n)$ is chosen as 
\begin{equation}\label{3-9}
f(n)=q^{n(n-1)/4} \ , \quad {\hbox{\rm i.e.,}}\quad
f(n)=q^{(n-1)/2} f(n-1) \ .
\end{equation}
As was stressed by them, if $0<q\le 1$, the infinite 
series (\ref{2-7b}) is convergent for any value of $|\gamma|^2$. 
For $f(n)$ given in Eq.(\ref{3-9}), we have $f(0)=f(1)=1$ and 
for $n=2, 3, 4, \cdots$, the same result as shown in Eq.(\ref{3-5}). 
Therefore, we obtain the same forms as those shown in Eq.(\ref{3-6}) 
and the state $\ket{c}$ in the general case is also 
in the intermediate mixture between the above two cases. 
However, the mechanism of the mixture may be different from the 
most popular form. 
In this case, $[n]_q$ and $[n]_q!$ are given by 
\begin{eqnarray}
& &[n]_q= nq^{-(n-1)} \ , \label{3-10}\\
& &[n]_q!=q^{-n(n-1)/2} n! \ .\label{3-11}
\end{eqnarray}
For $n=0, 1$, the relation (\ref{3-11}) gives us 
$[0]_q!=[1]_q!=1$. 
If $F({\hat N})=0$, the relation (\ref{2-22}) is reduced to 
\begin{equation}\label{3-12}
{\hat \gamma}{\hat \gamma}^*-q^{-1}{\hat \gamma}^*{\hat \gamma}
=q^{-{\hat N}} \ .
\end{equation}
The above is different from the previous form (\ref{3-7}). 

The above case can be easily extended by putting $f(n)$ in the form 
\begin{eqnarray}
f(n)&=&q^{C(n,r)/2} \ , \qquad
C(n,r)=n!/(n-r)!r! \ , \qquad \hbox{\rm for} \quad n \ge r \ , 
\label{3-13}\\
f(n)&=&1 \ , \qquad \hbox{\rm for} \quad n=0, 1, 2, \cdots, r-1 \ .
\label{3-14}
\end{eqnarray}
For $n=r, r+1, \cdots$, $f(n)$ is reduced to 
\begin{subequations}\label{3-15}
\begin{eqnarray}
& &f(n)\rightarrow 0 \qquad{\hbox{\rm for}}\qquad 
q\rightarrow 0 \ , \label{3-15a}\\
& &f(n)=1 \qquad{\hbox{\rm for}}\qquad 
q=1 \ . \label{3-15b}
\end{eqnarray}
\end{subequations}
Therefore, for the above two cases, $\ket{c}$ can be expressed as 
\begin{subequations}\label{3-16}
\begin{eqnarray}
\ket{c}&\rightarrow& 
(\sqrt{\Gamma})^{-1}\sum_{n=0}^{r-1}(\sqrt{n!})^{-1}
\gamma^n\ket{n} \nonumber\\
&=&\left(\sum_{n=0}^{r-1}(n!)^{-1}(|\gamma|^2)^n\right)^{-1/2}
\cdot \sum_{n=0}^{r-1}(n!)^{-1}(\gamma{\hat c}^*)^n \ket{0} \quad
{\hbox{\rm for}} \quad q\rightarrow 0 \ , \label{3-16a}\\
\ket{c}&=& 
(\sqrt{\Gamma})^{-1}\sum_{n=0}^{\infty}(\sqrt{n!})^{-1}
\gamma^n\ket{n} \nonumber\\
&=&\exp(-|\gamma|^2/2)\cdot \exp(\gamma{\hat c}^*)\ket{0} 
\ (=\ket{c^0}) \quad
{\hbox{\rm for}} \quad q=1 \ . \label{3-16b}
\end{eqnarray}
\end{subequations}
We can see that the form (\ref{3-16}) is extended from that given 
in Eq.(\ref{3-6}).

\subsection{Modified forms}

First, we treat the following two forms : 
\begin{eqnarray}
& &f(n)=\left[ 1-\exp(-C(n,r)^{-1}q(1-q)^{-1})\right]^{1/2} \ , 
\label{3-17}\\
& &f(n)=\sqrt{2}\left[ 1+\exp(C(n,r)q^{-1}(1-q))\right]^{-1/2} \ , 
\label{3-18}
\end{eqnarray}
Here, $C(n,r)$ is defined in Eq.(\ref{3-13}). 
The parameter $q$ runs in the domain $0<q\le 1$. 
For the above two forms, we can show the following relation : 
\begin{equation}\label{3-19}
f(n)=1 \qquad {\hbox{\rm for}}\qquad n=0, 1, 2, \cdots, r-1 \ .
\end{equation}
Further, for $n=r, r+1, \cdots$, $f(n)$ is reduced to 
\begin{subequations}\label{3-20}
\begin{eqnarray}
& &f(n)\rightarrow 0 \qquad{\hbox{\rm for}}\qquad q\rightarrow 0 \ , 
\label{3-209a}\\
& &f(n)=1 \qquad{\hbox{\rm for}}\qquad q=1 \ . 
\label{3-20b}
\end{eqnarray}
\end{subequations}
The relations (\ref{3-19}) and (\ref{3-20}) are exactly the same 
as those shown in Eqs.(\ref{3-14}) and (\ref{3-15}). 
Therefore, for the cases (\ref{3-17}) and (\ref{3-18}), 
we have the form (\ref{3-16}). 
There exists an infinite number of possible deformations. 
The forms (\ref{3-17}) and (\ref{3-18}) may be two of the possibilities. 
If $r=2$, the forms correspond to those shown in the 
relations (\ref{3-4}) and (\ref{3-9}). 

Next modification is found in the product of various $f(n)$ 
appearing already at some places. 
An example is the product of two $f(n)$ shown in the 
relations (\ref{3-4}) and (\ref{3-9}). In this case, the 
state $\ket{c}$ is in the intermediate mixture between 
the cases (\ref{3-6a}) and (\ref{3-6b}). 
The relations (\ref{3-3}) and (\ref{3-9}) give us 
the following relation for the producted new $f(n)$ : 
\begin{equation}\label{3-21}
f(n)=\sqrt{n(1-q^{-2})/(1-q^{-2n})}\ f(n-1) \ . 
\end{equation}
By substituting the relation (\ref{3-21}) into 
$[n]_q$ shown in Eq.(\ref{2-19}), $[n]_q$ is obtained in the form 
\begin{equation}\label{3-22}
[n]_q=(1-q^{-2n})/(1-q^{-2}) \ .
\end{equation}
The deformation deduced in the last equation has been 
studied before in Ref.\citen{AC} for $Q=q^2$. 
Then, the relations (\ref{2-15}) in this case become 
\begin{subequations}\label{3-23}
\begin{eqnarray}
& &{\hat \gamma}^*{\hat \gamma}=[{\hat N}]_q
=(1-q^{-2{\hat N}})/(1-q^{-2}) \ , 
\label{3-23a}\\
& &{\hat \gamma}{\hat \gamma}^*=[{\hat N}+1]_q
=(1-q^{-2({\hat N}+1)})/(1-q^{-2}) \ . 
\label{3-23b}
\end{eqnarray}
\end{subequations}
The above relations give us 
\begin{equation}\label{3-24}
{\hat \gamma}{\hat \gamma}^*-q^{-2}{\hat \gamma}^*{\hat \gamma}
=1 \ .
\end{equation}
The above should be compared with the relations (\ref{3-7}) and 
(\ref{3-12}). 
The form is quite simple. 

Third is related with the product of $f(n)$ appearing 
already and new function $g(n)$, i.e., 
$f_M(n)=f(n)g(n)$. 
Of course, $g(n)$ obeys the same condition as that shown 
in the relation (\ref{2-7c}). 
As $f(n)$, we can adopt, for example, $f(n)$ shown in the relation 
(\ref{3-21}). 
Then, the state $\ket{c}$ is expressed as 
\begin{equation}\label{3-25}
\ket{c}=(\sqrt{\Gamma})^{-1}
\sum_{n=0}^{\infty} f_M(n) (\sqrt{n!})^{-1} \gamma^n \ket{n} \ .
\end{equation}
The state $\ket{c}$ at the limit $q\rightarrow 0$ and 
the case $q=1$ can be given as 
\begin{subequations}
\begin{eqnarray}
& &\ket{c}=(1+|\gamma|^2)^{-1/2} 
(1+\gamma{\hat c}^*)\ket{0} 
\qquad{\hbox{\rm for}}\quad q\rightarrow 0 \ , 
\label{3-26a}\\
& &\ket{c}=\left(\sqrt{{\Gamma}}\right)^{-1} 
\sum_{n=0}^\infty g(n)(\sqrt{n!})^{-1} \gamma^n \ket{n} \ , \nonumber\\
& &\ \ {\Gamma}=\sum_{n=0}^\infty g(n)^2 (n!)^{-1}
(|\gamma|^2)^n \qquad {\hbox{\rm for}}\quad q=1 \ . 
\label{3-26b}
\end{eqnarray}
\end{subequations}
It is important to see that for $q=1$, the state $\ket{c}$ is 
not the conventional boson coherent state.

\section{Multiboson states}

One of the motivations 
of the present work comes from the investigation 
of a problem how to describe time-evolution 
of multiboson coherent state in terms of the time-dependent 
variational method.\cite{KPP} 
First, we prepare another space constructed by 
the boson operator $({\tilde c}, {\tilde c}^*)$, which is 
independent of $({\hat c}, {\hat c}^*)$. 
In this boson space, we introduce the following state : 
\begin{subequations}\label{4-1}
\begin{eqnarray}
& &\kket{c}=\left(\sqrt{\Gamma}\right)^{-1}
f_m({\tilde N})\exp(\sqrt[m]{\gamma}{\tilde c}^*)\kket{0}\ , 
\label{4-1a}\\
& &f_m({\tilde N})=(1/m)(1-e^{2\pi i{\tilde N}})/
(1-e^{2\pi i{\tilde N}/m}) \ , \quad
({\tilde N}={\tilde c}^*{\tilde c}) 
\label{4-1b}
\end{eqnarray}
\end{subequations}
\begin{equation}\label{4-2}
m=2, 3, 4, \cdots \ . \qquad\qquad\qquad\qquad \qquad\qquad\qquad\qquad
\end{equation}
It is not necessary to interpret the meaning of the 
notations. 
The state $\kket{c}$ can be rewritten as 
\begin{eqnarray}
& &\kket{c}=\left(\sqrt{\Gamma}\right)^{-1}
\sum_{n=0}^\infty \left(\sqrt{(mn)!}\right)^{-1}
\gamma^n\kket{mn} \ , \label{4-3}\\
& &\kket{mn}=\left(\sqrt{(mn)!}\right)^{-1} ({\tilde c}^*)^{mn}
\kket{0} \ , \label{4-4}\\
& &\Gamma=\sum_{n=0}^\infty \left((mn)!\right)^{-1}
(|\gamma|^2)^n\ .\label{4-5}
\end{eqnarray}
The operator $({\tilde c}^*)^m$ is the building block of the state 
$\kket{c}$ and we call the state $\kket{c}$ the multiboson 
coherent state. 
It may be self-evident that the infinite series (\ref{4-5}) 
is convergent in the domain $0\le |\gamma|^2 < \infty$. 

Structure of the state (\ref{4-3}) is different from that of 
the state (\ref{2-7a}) except $m=1$ and, then, we cannot apply the 
basic scheme presented in \S 2 to the state (\ref{4-3}) directly. 
In order to make it possible, we adopt the basic idea of the 
MYT boson mapping method\cite{MYT} ; the 
state $\ket{n}$ given in the relation (\ref{2-2a}) is the 
image of the state $\kket{mn}$ given in the relation 
(\ref{4-4}), i.e., we set up the correspondence 
\begin{equation}\label{4-6}
\kket{mn} \sim \ket{n} \ .
\end{equation}
The above correspondence permits us to introduce the mapping 
operator ${\mib U}$ in the form 
\begin{equation}\label{4-7}
{\mib U}=\sum_{n=0}^\infty \ket{n}\bbra{mn} \ .
\end{equation}
Then, the image of $\kket{c}$, which we denote $\ket{c}$, 
can be given in the form 
\begin{eqnarray}
& &\ket{c}={\mib U}\kket{c}
=\left(\sqrt{\Gamma}\right)^{-1}
\sum_{n=0}^\infty f(n) \left(\sqrt{n!}\right)^{-1}
\gamma^n\ket{n} \ , \label{4-8}\\
& &f(n)=\sqrt{n!}\left(\sqrt{(mn)!}\right)^{-1} \ .\label{4-9}
\end{eqnarray}
As is clear from the relations (\ref{4-8}) and (\ref{4-9}), 
we can apply the basic scheme to the state (\ref{4-8}). 
It may be interesting to see that the state (\ref{4-8}) can be 
rewritten as 
\begin{eqnarray}
& &\ket{c}=\left(\sqrt{\Gamma}\right)^{-1}
\exp\left(\left(\sqrt{m^m}\right)^{-1}\gamma{\hat c}^*
\left(\sqrt{F_m({\hat N})}\right)^{-1}\right)\ket{0} \ , \label{4-10}\\
& &F_m({\hat N})=\prod_{p=1}^{m-1} ({\hat N}+p/m) \ .
\label{4-11}
\end{eqnarray}
The images of ${\tilde c}^*{\tilde c}$ and $({\tilde c}^*)^m$, 
which we denote as $({\tilde c}^*{\tilde c})_c$ and $({\tilde c}^*)^m_c$, 
respectively, are expressed in the form 
\begin{subequations}\label{4-12}
\begin{eqnarray}
& &({\tilde c}^*{\tilde c})_c={\mib U}{\tilde c}^*{\tilde c}{\mib U}^{\dagger}
=m{\hat c}^*{\hat c}=m{\hat N} \ , 
\label{4-12a}\\
& &({\tilde c}^*)^m_c={\mib U}({\tilde c}^*)^m{\mib U}^{\dagger}
=\sqrt{m^m}{\hat c}^*\sqrt{F_m({\hat N})} \ . 
\label{4-12b}
\end{eqnarray}
\end{subequations}
With the use of $f(n)$ shown in Eq.(\ref{4-9}), we have 
\begin{equation}\label{4-13}
f({\hat N}+1)^{-1} f({\hat N})=\sqrt{m^m}\sqrt{F_m({\hat N})} \ .
\end{equation}
Then, ${\hat \gamma}$ and ${\hat \gamma}^*$ are given by 
\begin{equation}\label{4-14}
{\hat \gamma}=\sqrt{m^m}\sqrt{F_m({\hat N})}\ {\hat c} \ , \qquad
{\hat \gamma}^*=\sqrt{m^m}\ {\hat c}^*\sqrt{F_m({\hat N})} \ .
\end{equation}

Penson and Solomon have also investigated the multiboson coherent state 
in their framework. Let us discuss their form in the present scheme. 
Their state, which we denote $\kket{c}$, is expressed as 
\begin{eqnarray}
& &\kket{c}=\left(\sqrt{\Gamma}\right)^{-1}
\sum_{n=0}^\infty \sqrt{(mn)!}\ (n!)^{-1} q^{n(n-1)/4}
\gamma^n \kket{mn} \ , \label{4-15}\\
& &\Gamma=\sum_{n=0}^{\infty}
(mn)! (n!)^{-2} q^{n(n-1)/2} (|\gamma|^2)^n \ .
\label{4-16}
\end{eqnarray}
If $q<1$, the infinite series (\ref{4-16}) is convergent. 
But, if $q=1$, the series (\ref{4-16}) is convergent only for 
the case $m=2$. However, judging from the spirit of the 
present treatment, the state (\ref{4-8}) should be compared 
with the state (\ref{4-15}) at $q=1$. 
At this case, both are different from each other. 
Our state is normalizable, but theirs is not normalizable. 
However, we can introduce the effect of $q$ in terms of the product of 
$f(n)$ given in the relation (\ref{4-9}) and $f(n)$ discussed in 
\S 3, for example, such form as 
\begin{equation}\label{4-17}
f(n)=\sqrt{n!}\left(\sqrt{(mn)!}\right)^{-1} q^{n(n-1)/4} \ .
\end{equation}
Thus, for $0<q\le 1$, we have the state $\ket{c}$, which is 
normalizable and through the inverse process, we have $\kket{c}$, 
i.e., 
\begin{equation}\label{4-18}
\kket{c}={\mib U}^{\dagger} \ket{c} \ .
\end{equation}

\section{Classical aspect of the deformed boson scheme}

We return to the time-dependent variational equation 
(\ref{2-10}). In order to perform the variation, 
it is necessary to calculate 
the two quantities 
$\bra{c}i\partial_t \ket{c}$ and 
$\bra{c}{\hat H}\ket{c}$.\cite{KPTYs} 
Concerning the former, the time-derivative is obtained 
through the parameter $(\gamma, \gamma^*)$ and the result 
is as follows : 
\begin{equation}\label{5-1}
\bra{c}i\partial_t \ket{c}=(i/2)({\dot \gamma}\gamma^*
-{\dot \gamma}^*\gamma)\Gamma'/\Gamma \ .
\end{equation}
Here, $\Gamma$ is given in the relation (\ref{2-7b}) 
and $\Gamma'$ is given by 
\begin{equation}\label{5-2}
\Gamma'=d\Gamma/d|\gamma|^2
=\sum_{n=0}^\infty f(n+1)^2({n!})^{-1}(|\gamma|^2)^n \ge 0 \ .
\end{equation}
The expression (\ref{5-1}) can be changed in the form 
\begin{equation}\label{5-3}
\bra{c}i\partial_t \ket{c}=(i/2)({\dot z}z^*
-{\dot z}^* z) \ .
\end{equation}
The quantity $(z , z^*)$ is defined as 
\begin{equation}\label{5-4}
z=\gamma\sqrt{\Gamma'/\Gamma} \ , \qquad
z^*=\gamma^*\sqrt{\Gamma'/\Gamma} \ .
\end{equation}
The expectation value $\bra{c}{\hat H}\ket{c}$ is obtained as a 
function of $(\gamma , \gamma^*)$, i.e., 
$(z , z^*)$ : 
\begin{equation}\label{5-5}
\bra{c}{\hat H}\ket{c}=H \ .
\end{equation}
The variation (\ref{2-10}) gives us the following equation : 
\begin{equation}\label{5-6}
i{\dot z}={\partial H}/{\partial z^*} \ , \qquad
i{\dot z}^*=-{\partial H}/{\partial z} \ .
\end{equation}
The above is nothing but the Hamilton's equation of motion. 
The quantity $(z, z^*)$ can be regarded as 
the canonical variable in boson type. 
The above formulation is in parallel with the TDHF theory 
in canonical form. By solving Eq.(\ref{5-6}) under an appropriate 
initial condition, we can determine the time-dependence 
of $(z, z^*)$, i.e., $(\gamma, \gamma^*)$. 
Then, we have the state $\ket{c}$ as a function of time. 
It is noted that the above process is possible to perform 
if $(\gamma, \gamma^*)$ is expressed in terms of $(z, z^*)$ 
in the relation (\ref{5-4}). However, in general, it may be 
impossible 
to get an 
analytical form. Then, we will try to 
describe the system under investigation in terms of 
$(\gamma, \gamma^*)$. For this aim, we must investigate the physical meaning 
of the variable $(\gamma, \gamma^*)$. 

For the above investigation, first, we note the relation 
(\ref{2-13}), from which we have 
\begin{eqnarray}
& &\bra{c}{\hat \gamma}\ket{c}=\gamma \ , \qquad
\bra{c}{\hat \gamma}^*\ket{c}=\gamma^* \ , \label{5-7}\\
& &\bra{c}{\hat \gamma}^*{\hat \gamma}\ket{c}=\gamma^*\gamma \ .
\label{5-8}
\end{eqnarray}
Further, $\bra{c}{\hat N}\ket{c}$ is given as 
\begin{equation}\label{5-9}
\bra{c}{\hat N}\ket{c}=\gamma^*\gamma \Gamma'/\Gamma
=z^*z \ (=N) \ .
\end{equation}
With the use of the relation (\ref{5-4}), we can prove the 
relation : 
\begin{eqnarray}
& &[ N , \gamma ]_P=-\gamma \ , \qquad
[ N , \gamma^* ]_P=\gamma^* \ , \label{5-10}\\
& &[ \gamma , \gamma^* ]_P= d_N(\gamma^*\gamma) \ .\label{5-11}
\end{eqnarray}
Here, $d_N$ denotes the differential with respect to $N$ 
and $[ A , B ]_P$ expresses the Poisson bracket\footnote{ 
The definition of the Poisson bracket may be usually adopted 
as 
$\{ A, B\}_P=\partial A/\partial z\cdot \partial B/\partial (iz^*)
-\partial A/\partial (iz^*)\cdot \partial B/\partial z$. 
The relation to our definition (\ref{5-12}) is simply obtained 
as 
$i\{A, B\}_P=[ A, B ]_P$.
}
\begin{equation}\label{5-12}
[ A , B ]_P=\partial A/\partial z \cdot \partial B/\partial z^*
-\partial A/\partial z^*\cdot \partial B/\partial z \ .
\end{equation}
The relations (\ref{5-10}) and (\ref{5-11}) should be compared with 
the relations (\ref{2-17}) and (\ref{2-21}), together with 
the relations (\ref{5-7}), (\ref{5-8}) and (\ref{5-9}). 
Through the relation between the commutator and the Poisson bracket 
and, further, through the relation between the difference 
and the differential (in old quantum theory), 
we can conclude that the form presented in this section 
is a classical counterpart of the deformed boson scheme 
presented in \S 2 and we have the correspondence 
\begin{equation}\label{5-13}
{\hat \gamma}\sim \gamma \ , \quad 
{\hat \gamma}^*\sim\gamma^* \ , \quad
{\hat N}\sim N \ .
\end{equation}
Then, we can say that $(\gamma, \gamma^*)$ is a classical counterpart 
of the deformed boson $({\hat \gamma} , {\hat \gamma}^*)$.

The time-evolution of the classical counterpart of the 
deformed boson and/or deformed $su(2)$-generators has 
been investigated in the context of the $q$-deformed 
Lipkin model.\cite{CP} 
In our case, the Hamilton's equation of motion (\ref{5-6}) can be 
rewritten in terms of $(\gamma, \gamma^*)$ as follows :
\begin{eqnarray}
& &i{\dot \gamma}=\frac{\partial H}{\partial \gamma^*}
\left(\frac{\partial N}{\partial |\gamma|^2}\right)^{-1} \ , 
\qquad
i{\dot \gamma}^*=-\frac{\partial H}{\partial \gamma}
\left(\frac{\partial N}{\partial |\gamma|^2}\right)^{-1} \ , 
\label{5-14}\\
& &\frac{\partial N}{\partial |\gamma|^2}=
\frac{\Gamma'}{\Gamma}+|\gamma|^2\left(
\frac{\Gamma'}{\Gamma}\right)' \ . 
\label{5-15}
\end{eqnarray}
If $(\Gamma'/\Gamma)'=0$, the equation of motion 
(\ref{5-14}) is reduced to the conventional Hamilton's equation of 
motion. Solution of Eq.(\ref{5-14}) gives us the time-dependence 
of $(\gamma , \gamma^*)$ and, then, we obtain the state 
$\ket{c}$ as a function of $t$. For the calculation of 
$H \ (=\bra{c}{\hat H}\ket{c})$, it is necessary to give 
the expectation value of $({\hat c}^*)^r({\hat c}^*)^s({\hat c})^s$ 
and its hermit conjugate. 
The expectation value is given in the form 
\begin{equation}\label{5-16}
\bra{c}({\hat c}^*)^r({\hat c}^*)^s({\hat c})^s \ket{c}
=(\gamma^*)^r(|\gamma|^2)^s\cdot \Gamma_r^{(s)}/\Gamma \ .
\end{equation}
Here, $\Gamma_r^{(s)}$ is defined by 
\begin{equation}\label{5-17}
\Gamma_r^{(s)}=\sum_{n=0}^\infty 
f(n+r+s)f(n+s)({n!})^{-1}(|\gamma|^2)^n \ .
\end{equation}
In the case $r=0$, $\Gamma_0^{(s)}$ is given as 
\begin{eqnarray}\label{5-18}
\Gamma_0^{(s)}&=&\sum_{n=0}^\infty f(n+s)^2
({n!})^{-1}(|\gamma|^2)^n \nonumber\\
&=&(d/d|\gamma|^2)^s \Gamma = \Gamma^{(s)} \ .
\end{eqnarray}
In the case $r>0$, it is impossible to give the compact expression 
for $\Gamma_r^{(s)}$ and its derivative such as shown in the relation 
(\ref{5-18}). 
But, with the use of Cauchy (Schwarz) inequality, 
$\Gamma_r^{(s)}$ obeys 
\begin{equation}\label{5-19}
\Gamma_r^{(s)} \le \sqrt{\Gamma^{(r+s)}\Gamma^{(s)}} \ .
\end{equation}
Therefore, if the trial state $\ket{c}$ is not so different from 
the boson coherent state, the relation (\ref{5-19}) may be 
used as the equality.

\section{Concluding remarks}

In this paper, we presented our basic idea for the deformed boson scheme 
in the case of one kind of boson operator. This idea suggests us that 
the time-dependent variational method is also workable 
in terms of the variables which correspond to the deformed bosons. 
However, the deformed boson scheme displays its real ability 
in the case of two kinds of boson operators. 
In this case, we know two algebras : the $su(2)$- and the 
$su(1,1)$-algebra represented in terms of the Schwinger 
boson representation.\cite{Schw} 
In such systems, we investigated their time-evolutions in the 
framework of the coherent and the squeezed state.\cite{KPTYs} 
For this description, it 
would be 
interesting to give the deformed 
algebraic foundation 
because our treatment is systematic and general, 
although there is already work done in this topic  such as the work about $su(2)$-algebra 
in the Schwinger boson representation.\cite{D} 
This is our next subject, which will be 
discussed in (II).

\section*{Acknowledgements}

One of the present authors (Y. T.) would like to express 
his sincere thanks to Professor K. A. Penson for introducing 
his work about new generalized coherent states and giving him 
a short lecture when he stayed in the Laboratoire de 
Physique Th\'eorique des Particules El\'ementaires (LPTPE) 
of Universit\'e Pierre et Marie Curie (Paris VI). 
He also expresses his sincere thanks to Professor 
Dominique Vautherin for giving him a chance to discuss 
this problem and his great encouragement.

\end{document}